# Leveraging on Easy Java Simulation tool and open source computer simulation library to create interactive digital media for mass customization of high school physics curriculum

Loo Kang WEE

Ministry of Education,
Education Technology Division
Singapore, Singapore
weelookang@gmail.com

Wai Keong MAK

Centre for Applied Research
SiM University
Singapore, Singapore
wkmak@unisim.edu.sg

**This paper highlights the diverse possibilities in the rich community of educators from the Conceptual Learning of Science (CoLoS) and Open Source Physics (OSP) movement to engage, enable and empower educators and students, to create interactive digital media through computer modeling. This concept revolves around a paradigmatic shift towards participatory learning through immersive computer modeling, as opposed to using technology for information transmission.**

**We aim to engage high school educators to professionally develop themselves by creating and customizing simulations possible through Easy Java Simulation (Ejs) and its learning community. Ejs allows educators to be designers of the simulation. Educators can conduct lessons with students' using these interactive digital simulations and rapidly enhance the simulation through changing the source codes personally.**

**Ejs toolkit, its library of simulations and growing community contributed simulation codes, in a Web 2.0 environment, potentially allows for rapid proliferation of mass customized virtual laboratories simulation with educators as designers of learning environment.**

**Our journey of remixing this simulation created an educator-customized interactive digital media virtual laboratory for advancing learning physics by inquiry.**

**The process of creating a simulation also has desirable pedagogical value that will be challenging to bring to high school students to learn by modeling. The impediments to using this method are discussed.**

*Keywords- easy java simulation physics education teacher professional development integrated e-learning applet learning community open source self directed learning collaborative ICT masterplan 21st century skills modeling teach less learn more thinking school nation*

## I. Introduction

In the continued efforts to enrich and transform the learning environment in and outside formal school education, we seek to highlight the community of educators in the Conceptual Learning of Science (CoLoS) [1] and Open Source Physics (OSP) [2] movement to engage, enable and empower educators and students, to create interactive digital media through computer modeling.

The concept revolves around a paradigmatic shift towards participatory learning for deep and immersive computer modeling, as opposed to using technology for transmission of information.

We recognize the prerequisite knowledge and educational pedagogy necessary for such learning is suited for undergraduate learners, attempts to bring this technology enabled pedagogy to high school students, is exciting.

On a more pragmatic note, we will be sharing with educators, existing library of ready made Easy Java Simulation (Ejs) [3] simulations and ways to customize simulations to suit the needs of their students.

This paper highlights a legal and free way of enabling educators and students to learn and modify existing source codes to achieve finer customization of interactive digital media for enhancing learning for physics curriculum.

The Easy Java Simulations is created by, Francisco Esquembre who distributes it under a GNU GPL license [4] allows Ejs and its JAR library files to be copied and distributed freely.

Authors can distribute any simulation that they create and remix using Ejs, preserve original authors and contributors of the source codes and link to the official Web page for Ejs: http://fem.um.es/Ejs.

In addition, educators like Fu-Kwun Hwang licensed his source codes under Creative Commons Attribution 2.5 Taiwan License [5], which means anyone is free to share and remix the source codes under the condition of attribution of original author.

These innovative licenses aim to lower barriers to collaborate in today's media rich and highly interconnected world that we live in. Free and open source education resources are necessary nurturing conditions for spreading of resources, and co-creation and co-authorship of learning resources that can develop educators as designers of learning environments. Now, educators and students can stand on the shoulders of giants and learn about simulation building by tinkering creative commons license source codes.

## II. Installation of Easy Java Simulations and learning activities

New readers to Easy Java Simulations (Ejs) may want to refer to the website to download the latest version of Ejs and Java Runtime Environment (JRE). Additional materials like Modeling Science Textbook Chapter 2: Introduction to Easy Java Simulations [6], Modeling Science Textbook Chapter 3: Ejs and Java Concepts [7] and Modeling Physics with Easy Java Simulations [8] are valuable starter learning activities.

### III. MOTIVATION FOR USING EJS

The concepts in linear momentum and kinetic energy and its conservation in perfectly elastic and inelastic collisions are difficult concepts for many A- level Physics students. Ejs allows educators to be designers of learning environments to allow for inquiry learning. A customized interactive digital learning environment can be a manipulative tool used by students.

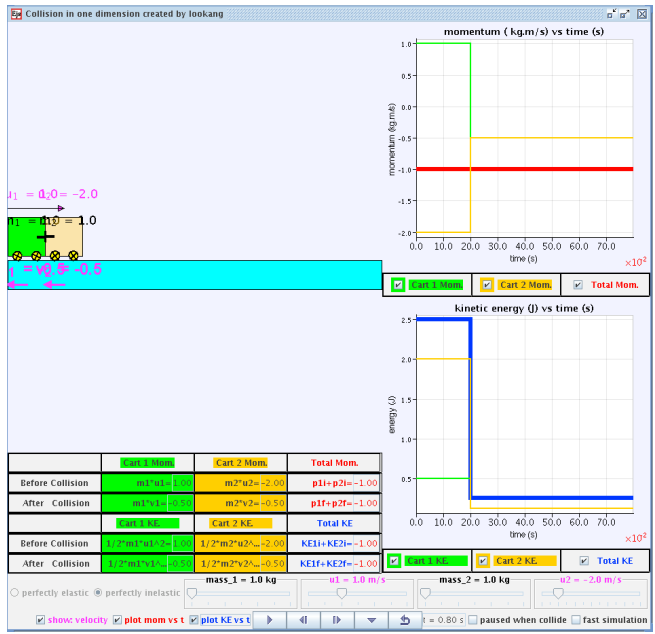

Figure 1.   Ejs Applet View of the interative digital media and virtual laboratory.

Educators can pilot the use of these interactive digital simulation, listen to the users' feedback and improve the interactive digital simulation without always relying on difficult enhancement processes such as paying vendors to build and add modifications later. Ejs creative commons and open source codes on Web 2.0 platforms, allow nurturing of ecology of educators to rapidly proliferate of such interactive digital simulation. Mass customizing of learning digital media, both the process as well as the interactive digital simulation itself, can serve to work towards realizing MOE's vision of "Thinking School, Learning Nation" TSLN, engage learners to prepare for life through Teach Less, Learn More (TLLM) - transforming learning [9].

In addition, Ejs also presents a possible unified approach in the building virtual laboratories [10], a standardized platform for creating interactive digital media, allowing continual improvements by anyone.

### IV. LEARNING JOURNEY OF MODIFYING SIMULATIONS

Easy Java Simulations (Ejs) is an authoring tool developed for the conceptual learning of science. As high school educators, we are able to design, create or modify scientific simulations, especially with the examples in Ejs itself, CompADRE Digital Library [11] and Web 2.0 community online forum like NTNUJAVA Virtual Physics Laboratory [12]. We can concentrate our effort in writing and refining the relations in the underlying scientific model. Rapid prototyping allows us to improve learners' user experience, build in scaffolds and hints for sense making. Through modifying or creating simulations, we are engaged in constructive modeling, improving our understanding of computational physics in this computer enabled pedagogy [13]. Much of our journey requires us to be self directed as well as to work collaboratively in the Ejs learning community. We strengthen our ability to integrate ICT meaningfully by creating inquiry learning environments for sense making, aligned with Third Masterplan for ICT in Education [14].

### V. ONE DIMENSIONAL COLLISION VIRTUAL LABORATORY DESIGN FOR A-LEVEL PHYSICS TOPIC: LINEAR MOMENTUM AND ITS CONSERVATION

In our exploration on Ejs as a tool for teaching and learning, we have already create a few examples of remixing and customizing existing source codes into useful interactive digital media [15].

After Ejs is downloaded from the Ejs website, unzip it into a directory of your choice, for example D:\Shared\EasyJavaSimulation.

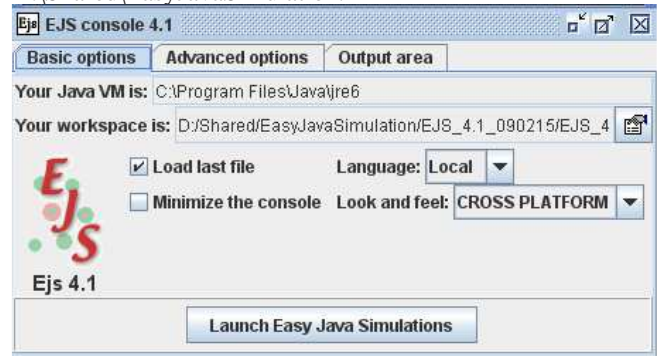

Figure 2.   Example of Ejs console with options setting.

Executing the EjsConsole.jar file in the installation directory will launch the console and the building interface.

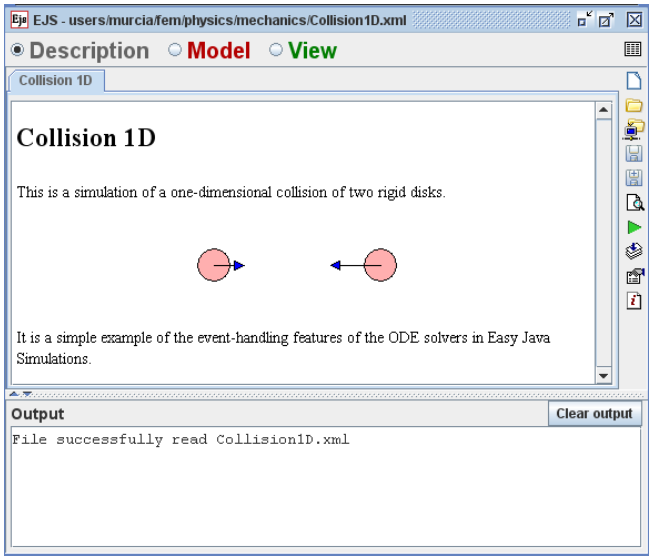

Figure 3. Example of Ejs interface.

Open the file found at **D:\Shared**\EasyJavaSimulation\Ejs_4.1_090215\Ejs_4.1\workspace\source\users\murcia\fem\physics\mechanics\Collision1D.xml depending on where the installation is unzipped. It is a suitable source file for remixing into a virtual laboratory environment for A-level physics. This is one of the many simple models made for teaching Ejs as a modeling tool and as an introduction to Ejs open programming capability.

In our paper, we assume the reader has some knowledge about Ejs so refer to [6], [7] and [8] for other starter activities. After examining the variables, evolution, simulation view, analyzing and tinkering with the codes, to get an idea of how to model works, we pose ourselves the question: "How can we build on this existing expert source codes?" Next, we conceptualize how this existing model can be remixed into a virtual laboratory learning environment with us, the educators as designers of learning environments.

Use the "Save As" icon on the Ejs taskbar to create a copy of this file in your own working directory, for example Ejs_4.1_090215\Ejs_4.1\workspace\source\users\your_organisation\your_name\Collision1Dyour_name.xml.

Designing a virtual laboratory environment is a non-linear process but we will discuss in brief some of the key stages.

## A. Look and Feel

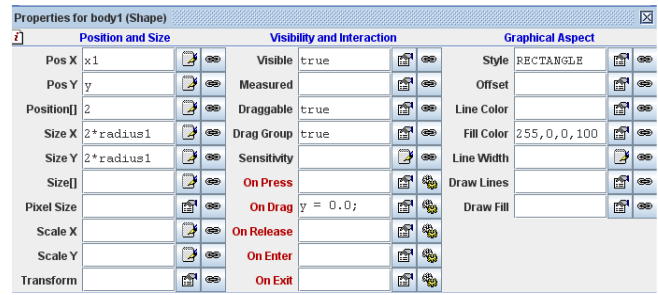

Figure 4. Example of Ejs Property for Shape body1 with variables inputs.

We visualize the simulation space to mimic real life collision carts setup. By exploring the existing property of body1, the concepts of position x and y, size of the object in x and y direction, style and fill color, we can begin to tinker and change the visual property of cart 1.

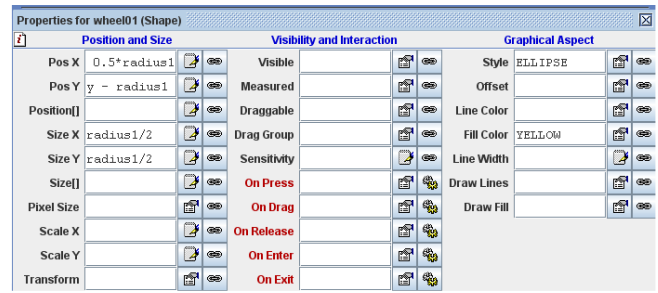

Figure 5. Example of Ejs Property console with variables inputs for Shape wheel01 with variables inputs..

By adding the wheels to the carts, we learn about relative position in order for the position of the wheel to be in relation to the variables x1 and y. It is now possible to group the items under a 2D group and input the position parts of the cart1 referencing that 2D group relative positions.

By shifting the position in the tree of elements, we learn about the role of foreground and background like layers concept in most drawing tools, in order to achieve the look of the carts with the wheels in front of the cart instead of behind it.

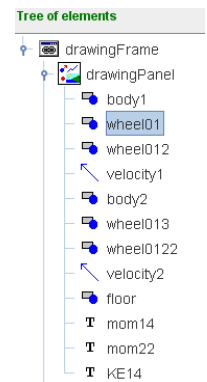

Figure 6. Example of Ejs Tree of elements in VIEW for drawing Panel.

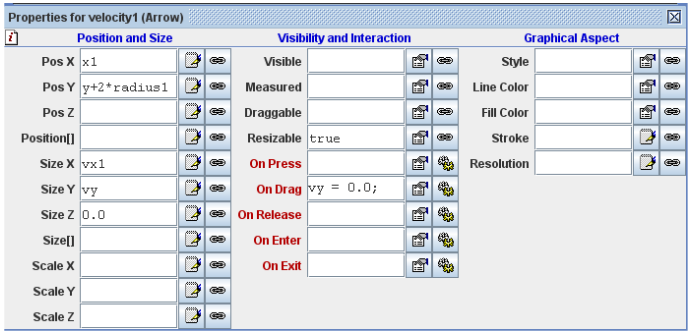

Figure 7. Example of Ejs Property for Arrow velocity1 with variables inputs.

Based on our students' previous mental representation of how velocity is visualized, we adjust the Pos Y equal to y+2*radius1 to be above the cart. This allows students to quickly relate what they see in textbooks to what they interactive with in the virtual laboratory.

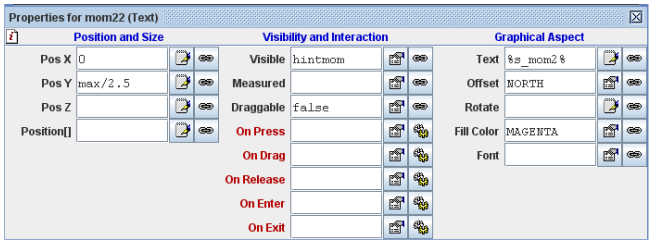

Figure 8. Example of Ejs Property for Text mom22 with variables inputs.

We design pedagogical hints into the simulation view by determining the position x and y for it to appear, the visible when hintmom (Boolean variable) is true, the text to appear %s_mom2% , offset in the position NORTH if needed and fill color set to magenta.

The steps may seem very programming-like but the codes are available in source codes shared by the community. We just need to copy it with accreditation and adapt to our needs accordingly.

CODE I.          DISPLAY NUMBER IN CORRECT FORMAT

```
// this code has to do with displaying the number 2.3
//instead of 2.300000001 when using % %
// donated by Fu-Kwun Hwang
double dv;
public String double2String(double v,double divider){
  if(v>0)dv=0.5; // if v>0  2.3 become 2,  2.5 become 3
  else if(v<0)dv=-0.5; // -2.3 become -2, -2.5 become -3
  else dv=0;
  return (int)(v*divider+dv)/divider+"";
}
```

● Fixed relations Page:

CODE II.          DISPLAY OF STRING MOMENTUM EQUATION

```
s_mom2 = double2String(mass1,1.)+" * (
"+double2String(v_temp1,1000.)+" ) + "+
double2String(mass2,1.)+" *
("+double2String(v_temp2,1000.) +" ) = " +
double2String(mass1,1.)+" *
("+double2String(vxf1,1000.)+" ) + "+
double2String(mass2,1.)+" * (
"+double2String(vxf2,1000.)+ " ) ";
```

This allows a checkbox to be assigned ☑ hint: Conservation of Momentum the variable hintmom for displaying the dynamic hint for conservation of momentum that changes as the value varies:

$$m_1u_1 + m_2u_2 = m_1v_1 + m_2v_2$$

1.0 * ( 0.408 ) + 1.0 * (-0.401) = 1.0 * ( 0.0040 ) + 1.0 * ( 0.0040 )

The openness of the tool allows many possibilities for designing hints and scaffolds, perhaps only limited by the designer's coding knowledge and the Ejs community.

## B. Fixed Relations

We can declare more variables of the suitable type (Double, Integer, Boolean, String or Object) and input fixed relations such as these standard equations for calculation of the physics quantities.

CODE III.          EQUATIONS ADDED TO ALLOW FOR CALCULATIONS

```
//formula correct for 1-D motion
KE1 = 0.5*mass1*(vx1*vx1);
KE2 = 0.5*mass2*(vx2*vx2);
mom1 = mass1*vx1;
mom2 = mass2*vx2;
totalKE = KE1 +KE2 ;
totalmomentum = mom1 + mom2;
```

We start a ☐ panel for the simulation view and allow these variables to be shown. For example, start a panel on the left with layout VBOX and key in the label text (usually name of variable) and the symbol "=", the field variable defined and lastly a label of the units as text.

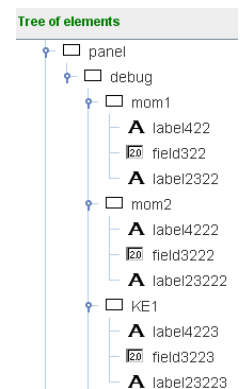

| | | |
|---|---|---|
| mom_1 = | 0.408 | kg.m/s |
| mom_2 = | -0.401 | kg.m/s |
| KE_1 = | 0.083 | J |

Figure 9.  Example of Ejs Tree of elements in VIEW for panel.

For ease of user interaction, 2 radio buttons
○ perfectly elastic  ○ perfectly inelastic are designed for the legal
state of the collision types.

CODE IV.     ELASTIC AND INELASTIC STATE

```
if ( elastic == true ) {
   inelastic = false;
   e = 1.0 ;
}
if ( inelastic == true ) {
   elastic = false;
   e = 0.0 ;
}
```

We exercise creative problem solving and thinking, apply
some simple logic and programming syntax in java to allow
only one state to be selected and assign the suitable values
coefficient of restitution, e.
Ejs 4.1 has a code wizard to facilitate code syntax
compliance.

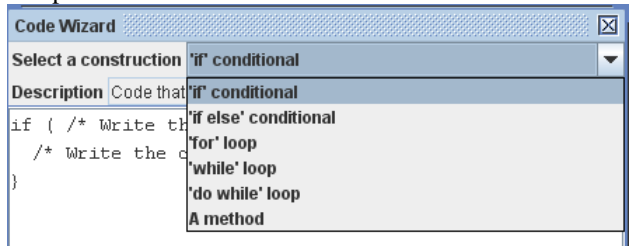

Figure 10.  Example of Ejs Code Wizard scaffolds.

## C.  Evolution

The equations of evolution are already in the sample source
codes provided in the Ejs download. They are defined in the
usual first order differential equations:

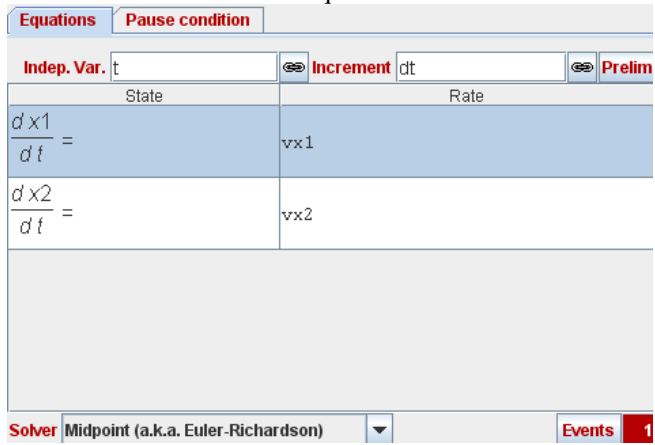

Figure 11.  Example of Ejs Equations View in Evolution Page.

Inside the Events button, an event is specified by providing:
the zero condition, the desired tolerance, and the action to
invoke.

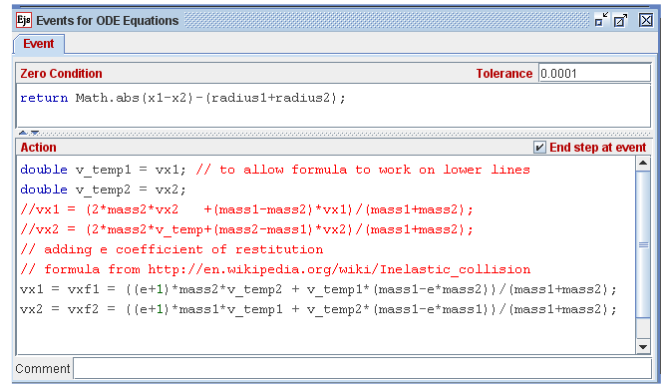

Figure 12.  Example of Ejs Event for Ordinary Differential Equations in
Evolution Page.

We learn to use the information on the internet [16] and
adapt to our coding variable assignment, engaging in
contextual problem solving.

$$v_1 = \frac{(e+1)m_2u_2 + u_1(m_1 - e.m_2)}{m_1 + m_2}$$

$$v_2 = \frac{(e+1)m_1u_1 + u_1(m_2 - e.m_1)}{m_1 + m_2}$$

Here, we apply what is typically not taught to A-level
physics students, but may help students learn for life-long
learning, apply to a programming environment for meaning
making and deep learning.
We had to look at the existing codes, make sense of it, and
device ways to implement the formula in physics textbooks
in the coding. The immediate feedback from the Ejs tool
allow us to reflect, tinker the codes, engage in deep problem
solving, communicate with community of experts in Web
2.0 forums and learn beyond.
In the pause condition tab, we can change the code to pause
the simulation.

CODE V.     PAUSE CONDITION

```
if (x1>(max-radius1 || x1<-max+radius1 || x2>max-
radius2 || x2<-max+radius1) _pause();
```

For advanced learners, it is possible to create a challenge
task for deep contextual learning by posing the question
"How can the carts pause exactly at the edge of the screen?"
Here, we learn to reposition the codes and even a way to
stepback the simulation by –dt so as to achieve the
appearance of pausing exactly at the edge.

These are learning experiences, made possible with
computers and Ejs.

## D.  Virtual Laboratory Design

For this learning environment to be used for inquiry
learning context, the simulation can be designed with simple sliders for variables with enough variation of values to allow for pattern recognition and trends to emerge.

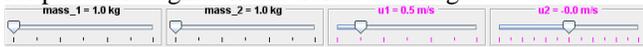

Figure 13. Ejs Slider control for variables in Applet.

The slider values can be selected to snap to closest, for ease of values selection 1, 2, 3, 4, 5, 6, 7 kg for example instead of 1.354 kg type of values.

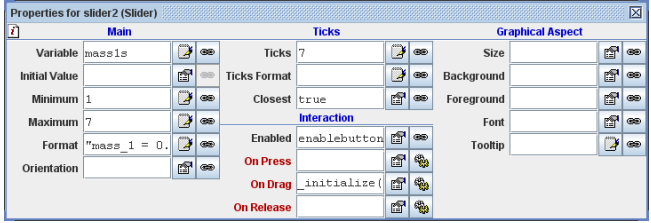

Figure 14. Example of Ejs Property for Slider slider2 with variables inputs.

The variable is assigned to mass1s, minimum value of 1 and maximum of 7, format is mass_1 = 0.0 kg (display will be 7.0, change to 0.00 if desired display is 7.00) , number of ticks is 7, closest is true, enabled is true when enablebutton (Boolean type) is true and on drag run function _initialize() ( to run the codes in the initialization page).

In the initialization page, we assign back to slider values in the variable mass1, for the rest of the calculations.

CODE VI.        MASS REMEMBER STORED VALUES

```
// add code to make mass remember stored values
mass1= mass1s ;
mass2= mass2s ;
```

### E. Dynamic graph plotting capability

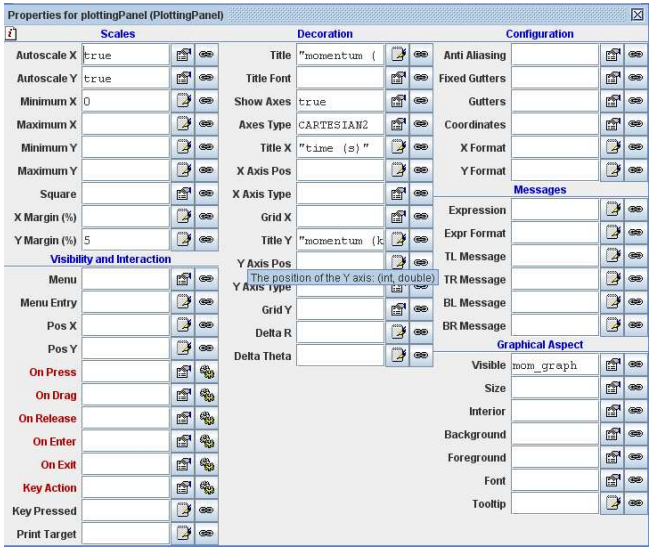

Figure 15. Ejs Property for Plotting Panel with variables inputs.

In the plotting panel, quantities for the plotting panel graph paper shown as defined and self-explanatory.

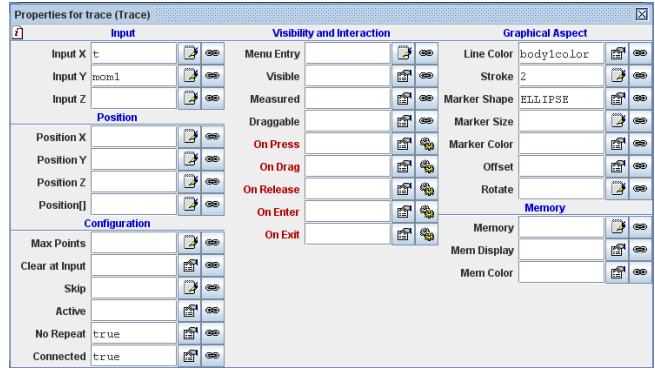

Figure 16. Ejs Property for trace with variables inputs.

The trace defined what is to be drawn. The input X is variable t, the input Y is mom1 for momentum of cart 1 depending on your assignment of variables, no repeat set to true, connected set to true (false will give the data plotted look like real data loggers), line color is assigned body1color variable object (for ease of changing color), stroke is 2 (thickness of line drawn) and marker shape is ellipse.

In fixed relations, this code is added for ease of changing color.

CODE VII.        ASSIGN VARIABLE BODY1COLOR

body1color = new java.awt.Color(0,255,0,255); //green is 0,255,0 with transparent 255

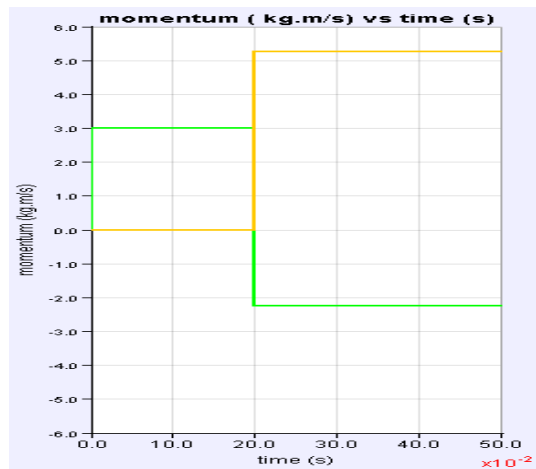

Figure 17. Ejs Plotting Panel Graph for momentum in applet.

We learn that Ejs is an easy tool to create visually accurate and dynamically linked graphs with axes of our choice.

## VI.    IMPEDIMENTS TO USING EJS AS PEDAGOGICAL TOOL

Established pen and paper education system and assessment makes it a challenge to create learning experiences through tinkering with the source codes, such as those possible with Ejs. There is also the issue of difficulty to learning just physics, because Ejs requires learning of physics content as well as other computer programming, problem solving and critical thinking skills.

Established school culture, teacher pedagogical understanding of meaningful learning, technology maturity and modeling skills are barriers to implementing these learning strategies.

Steep learning curve on both the educators and the students is demanded for learning to modify codes in this new technology constructionism pedagogy. Perhaps the beginning learning tasks need to be scaffold, progressing from very simple activities to more challenging ones.

## VII. CONCLUSION

It is today possible to engage in informal learning of changing or tinkering with codes and interact with the world experts in Ejs simulation building in Web 2.0 forum champion by Professor **Fu-Kwun Hwang.**

There is a growing number of educators using Ejs and sharing their source codes under creative commons licenses that allows others who are still beginning to use Ejs to create derived works. These experts usually share their codes that can form the basic algothrim and educators can later understand, improve and redesign with customization to their classroom needs. This approach is less taxing on educators coding skills, and presents a possible way for educators to leverage on open source Easy Java Simulation tool and computer simulation library to create interactive digital media for mass customization of high school physics curriculum, contributing back to global movement for open source and free education.

Efforts to use Ejs as a pedagogical technology tool are currently being implemented in the undergraduate physics education. Pushing the frontier of learning by construction and modeling presents many challenges to transform physics education but a meaningful step towards learning by doing and modeling.

We view Ejs as an open tool for learning suited for undergraduate physics education, a possible way for educators to customize learning virtual laboratory and simulations and an excellent modeling tool for high school students to learn problem solving skills and other critical competencies and dispositions.

Creating the learning experiences where students engage in self directed and collaborative learning with world experts through model creation, in formal and the informal spaces, may be a suitable proxy to the kind of learning they engage in later in their adult life.

The virtual laboratory is freely accessible on the internet as a standalone downloadable Ejs jar file and the source codes is available as downloadable Ejs XML source [17], complete with worksheets for learning by doing.

## ACKNOWLEDGMENT

We wish to acknowledge the passionate contributions of Francisco Esquembre and Fu-Kwun Hwang for their ideas, knowledge and insights in the co-creation of interactive digital media for education.